\documentclass[twocolumn,apj,appendixfloats,numberedappendix]{emulateapj-rtx4}

\usepackage{natbib}
\usepackage{graphicx}
\usepackage{amsmath,amssymb}
\usepackage{graphicx}
\usepackage{txfonts}
\usepackage{epstopdf}

\usepackage{ulem} 
\usepackage{natbib}
\newcommand{\vect}[1]{{\boldsymbol{#1}}}

\newcommand{\bnabla}{\boldsymbol{\nabla}}

\begin{document}
\title{Solar wind turbulence from MHD to sub-ion scales: high-resolution hybrid simulations}
\author{Luca~Franci\altaffilmark{1}}
\affil{Dipartimento di Fisica e Astronomia, Universit\`a di Firenze,
Largo E. Fermi 2, I-50125 Firenze, Italy.}
\author{Andrea~Verdini\altaffilmark{2}}
\affil{Dipartimento di Fisica e Astronomia, Universit\`a di Firenze,
Largo E. Fermi 2, I-50125 Firenze, Italy.}
\author{Lorenzo~Matteini}
\affil{Department of Physics, Imperial College London, London SW7 2AZ, UK.}
\author{Simone~Landi \altaffilmark{3}}
\affil{Dipartimento di Fisica e Astronomia, Universit\`a di Firenze,
Largo E. Fermi 2, I-50125 Firenze, Italy.}
\author{Petr~Hellinger}
\affil{Astronomical Institute, AS CR, Bocni II/1401,CZ-14100 Prague, Czech Republic}

\altaffiltext{1}{INFN - Sezione di Firenze, Via G. Sansone 1, I-50019 Sesto F.no (Firenze), Italy.}
\altaffiltext{2}{Solar-Terrestrial Center of Excellence - SIDC, Royal
Observatory of Belgium, Bruxelles, Belgium.}
\altaffiltext{3}{INAF, Osservatorio Astrofisico di Arcetri, Largo E. Fermi 5, I-50125 Firenze, Italy.}

 \date{\today}
 
\begin{abstract}
We present results from a high-resolution and large-scale hybrid (fluid electrons and particle-in-cell protons)
two-dimensional numerical simulation of decaying turbulence.
Two distinct spectral regions (separated by a smooth break at proton scales) develop with clear power-law scaling, each one occupying about a decade in wave numbers.
The simulation results exhibit simultaneously several properties of the observed solar wind fluctuations:
spectral indices of the magnetic, kinetic, and residual energy spectra in the magneto-hydrodynamic
(MHD) inertial range along with a flattening of the electric field spectrum,
an increase in magnetic compressibility, and a strong coupling of the cascade with the density and the parallel component of the
magnetic fluctuations at sub-proton scales. Our findings support the
interpretation that in the solar wind large-scale MHD fluctuations naturally
evolve beyond proton scales into a turbulent regime that is governed by the
generalized Ohm's law.
\end{abstract}
   \maketitle
\section{Introduction}
In-situ measurements of the solar wind plasma and electromagnetic field
show spectra with a power-law scaling spanning
several decades in frequency, $f$ \citep[e.g.][]{Alexandrova_al_2009, Sahraoui_al_2010, Roberts_2010}.
Power-laws support an interpretation in term of turbulent fluctuations, 
although the rich variety of spectral features is not easily explained in the
framework of known turbulent theories and phenomenologies.

For frequencies in the so-called magneto-hydrodynamic (MHD) range,
$10^{-4}~\mathrm{Hz}\lesssim f\lesssim 10^{-2}~\mathrm{Hz}$ at $1~\mathrm{AU}$,
the magnetic field spectrum and the kinetic field spectrum have a different
scaling, the former being proportional to $f^{-5/3}$ while the latter to $f^{-3/2}$
\citep{Podesta_al_2007,Salem_al_2009,Wicks_al_2011,Tessein_al_2009}. 
While a magnetic excess is generally found in solar wind turbulence, only recently the spectrum of residual energy (the difference between magnetic and kinetic energy) was shown to have a power-law scaling with a spectral index $-2$ \citep{Chen_al_2013b}.
Such finding confirms early predictions on the residual energy
spectrum \citep{Grappin_al_1983} and numerical results of
incompressible MHD simulations \citep{Muller_Grappin_2005}. 
Note that the three spectral indices ($-3/2,~-5/3,~-2$) for the kinetic, magnetic, and residual
energy spectrum are not reproduced simultaneously in any direct numerical
simulation (DNS) \cite[e.g.][]{Muller_Grappin_2005,Chen_al_2011a} unless a
particular driving is applied to large scales \citep{Boldyrev_al_2011}.
Finally, in the MHD range, magnetic and velocity fluctuations are dominated by the transverse components with respect to the ambient magnetic field $\vect{B}_0$ \citep[e.g.][]{Smith_al_2006b, Wicks_al_2011}.

Moving to higher frequencies, $f\gtrsim 10^{-2}~\mathrm{Hz}$,
there is growing evidence that kinetic effects become important and change the nature of the self-similar
spectra of fluctuations observed for $f\gtrsim1~\mathrm{Hz}$. 
A spectral break 
appears in magnetic and velocity spectra at proton scales, separating the MHD inertial range cascade from a second power-law interval at kinetic scales. The physical scale associated with this spectral break has not been identified yet \citep[e.g.][]{Bourouaine_al_2012,Bruno_Trenchi_2014, Chen_al_2014b}.
The spectral index of magnetic fluctuations after the break varies between
$(-4,-2)$ \citep{Leamon_al_1998, Smith_al_2006b}, although it tends to cluster around a slope
of $-2.8$ for higher frequencies \citep{Alexandrova_al_2012}. 
The change in the turbulence regimes also shows up in the density spectrum \citep{Chen_al_2013}, which steepens and couples to the parallel component of magnetic field. The latter becomes as energetic as the two
perpendicular components, resulting in an increase of the so-called magnetic
compressibility \citep{Alexandrova_al_2008b,Salem_al_2012,Kiyani_al_2013}.
Finally, measurements at 1 AU  show that the spectrum of electric field 
flattens at about $1~\mathrm{Hz}$ \citep{Bale_al_2005,Kellog_al_2006}, although
the noise level hinders the determination of a precise spectral scaling.

The measure of structure functions of third order 
at MHD scales \citep{SorrisoValvo_al_2007,MacBride_al_2008} 
and of high-order 
at MHD \citep{Salem_al_2009} and at sub-proton scales 
\citep{Kiyani_al_2013} provided additional evidences that 
fluctuations are turbulent all the way down to electron scales in the
solar wind.
While DNS are able to reproduce some aspects of
either the MHD range \citep[e.g.][]{Maron_Goldreich_2001, 
Mason_al_2008, Beresnyak_Lazarian_2009, Grappin_Muller_2010, Lee_al_2010, Chen_al_2011a, 
Boldyrev_al_2011,Dong_al_2014}
or the sub-proton range \citep[e.g.][]{Matthaeus_al_2008,Howes_al_2011,Markovskii_Vasquez_2011, Camporeale_Burgess_2011, Boldyrev_al_2013, Gary_al_2012, Wan_al_2012, Servidio_al_2012,Meyrand_Galtier_2013, Passot_al_2014}, 
to our knowledge a clear indication that a turbulent regime 
establishes in the whole spectrum spanning the two ranges has never been
reported so far.

In this work we present results from a high-resolution hybrid (fluid electrons,
particle-in-cell protons) two-dimensional (2D) DNS of turbulence and provide the first direct numerical evidence of the simultaneous occurrence of several features observed in the solar wind spectra.
These include i) the different scaling of magnetic and kinetic fluctuations in the MHD range, ii) a magnetic spectrum with a clear
double power-law scaling separated by a break, iii) an increase in
magnetic compressibility at small scales, iv) a strong coupling between
density and magnetic fluctuations at small scales.
The electric field spectrum is also consistent with observations, showing a
change in the spectral properties at sub-proton scales. 
Our results indicate that the switch in the spectral slopes 
observed in the solar wind results from the natural continuation of a large-scale MHD turbulent
cascade through proton and down to electron scales, where the different field couplings are governed by the non-ideal terms of the Ohm's law.
\section{Numerical setup}
The kinetic model uses the hybrid approximation: electrons are
considered as a massless, charge neutralizing, isothermal fluid;
ions are described by a particle-in-cell model 
(see \citealt{Matthews_1994} for detailed model equations).
The characteristic spatial and temporal units used in the model
are the proton inertial length $d_p=v_A/\Omega_{p}$, $v_A$ being the Alfv\'en speed,
 and the inverse proton gyrofrequency $1/\Omega_{p}$, respectively.
We use a spatial resolution
$\Delta x=\Delta y= 0.125 \, d_p$, and there are $8000$ particles-per-cell (ppc)
representing protons. The resistive coefficient is set to the 
value $\eta=5~10^{-4}~4\pi v_Ac^{-1}\Omega_p^{-1}$ to prevent the accumulation
of magnetic energy at the smallest scales.
Fields and moments are defined on a 2D $x$--$y$ grid with
dimensions $2048^2$ with periodic boundary conditions.
Protons are advanced with a time step $\Delta t=0.025 \, \Omega_p^{-1}$,
while the magnetic field $\boldsymbol{B}$ is advanced with a smaller time step $\Delta t_B = \Delta t/10$.
The number density $n$ is assumed to be equal for protons and electrons, $n_p=n_e=n$, and both protons and electrons are isotropic, with $\beta_p=\beta_e=0.5$ where $\beta_{p,e}=8\pi n K_B T_{p,e} / B^2_0$ are the proton (electron)
betas (here $K_B$ is the Boltzmann's constant, $B_0$ the ambient magnetic field, and $T_{p,e}$ are the proton and electron
temperatures). 

We impose an initial ambient magnetic field $\vect{B}_0 = B_0
\hat{\vect{z}}$, perpendicular to the
simulation plane. 
We add an initial spectrum of linearly
polarized magnetic and bulk velocity fluctuations $\vect{u}$, 
with only in-plane components. 
Fourier modes of equal amplitude and random phases are excited in the
range $-0.2 < k_{x,y} < 0.2$, assuring energy equipartition and vanishing
correlation between kinetic and magnetic fluctuations. Initial velocity
fluctuations have vanishing divergence and density fluctuations are also
vanishing (in the limit of numerical noise). 
Quantities are defined as parallel ($\|$) and perpendicular ($\perp$) with respect to $\vect{B}_0$.
We define the omnidirectional spectra,
\begin{eqnarray}
E_\Psi(k_\bot)\equiv\delta
\Psi^2(k_\bot)/k_\bot=\sum_{|\vect{k_\bot}|=k_\bot}\hat{\Psi}^2_{2D}(\vect{k}_\bot),
\end{eqnarray}
where $\hat{\Psi}$ are the Fourier coefficients of a given quantity $\Psi$ (we use
$\vect{E}$ and $\vect{J}$ to indicate electric field and current density
respectively) and $\delta \Psi(k_\bot)$ is the amplitude of the fluctuation
$\Psi$ at the scale $k_\bot$. We also define 
the root mean square value (rms) as
\begin{eqnarray}
 \Psi^{rms}=\sqrt{\langle \Psi^2\rangle - \langle \Psi
\rangle^2}.
\end{eqnarray}
where $\langle...\rangle$ stands for a real-space average over the whole simulation domain.
With these definitions, the initial conditions have 
$E_u\sim E_B\propto k_\bot$ with $B^{rms}/B_0\sim0.24$ allowing for a
fast turbulent dynamics sustained for about $300~\Omega_p^{-1}$  (the
nominal nonlinear time at the beginning of the simulation
is approximately $20~\Omega_p^{-1}$, but it increases at later time).

\section{Results}
\begin{figure}[t]
\begin{center}
\includegraphics [width=0.98\linewidth]{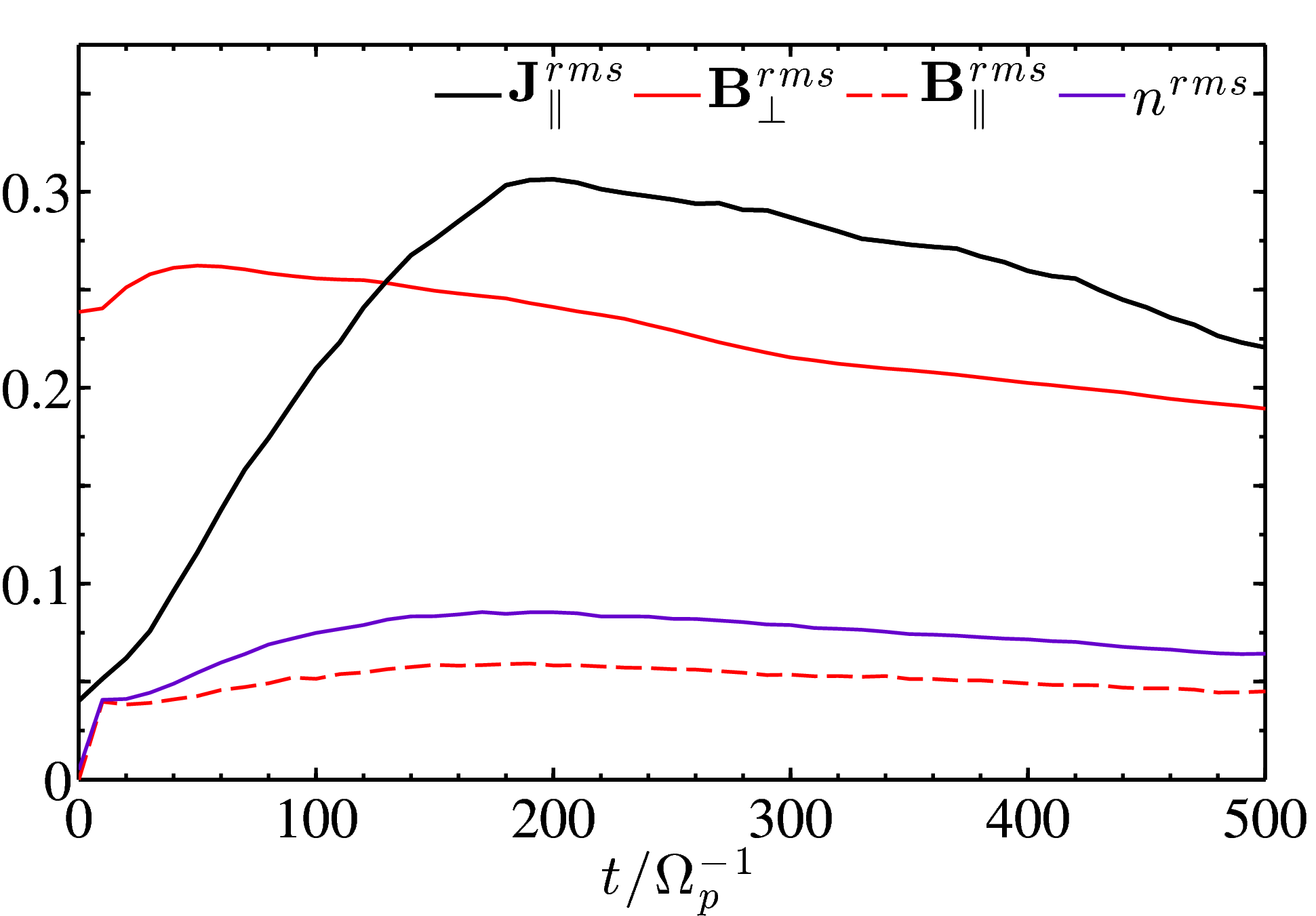}
\caption{Root-mean-square parallel current density (black line),
  perpendicular and parallel magnetic fluctuations (red-solid line and
  red-dashed lines respectively), and density fluctuations (purple
  line) as a function of time (normalized to the inverse of the proton
  gyrofrequency
  $\Omega_p$). 
  As a reference, the nonlinear time
  at the initial time is about $20~\Omega_p^{-1}$.}
\label{fig1}
\end{center}
\end{figure}
In Figure~\ref{fig1} we plot the rms of the parallel
current density, of the parallel and perpendicular magnetic field, and of the
density fluctuations. The current density increases until $t=200~\Omega_p^{-1}$, reflecting the formation of small scales due
to the development of a turbulent cascade, and then declines smoothly. The decay is slow, 
since larger and larger scales continue to feed the cascade at later times. Accordingly the perpendicular magnetic field declines steadily after a transient increase.
Shortly after the beginning, fluctuations in the parallel component of magnetic field and in the density appear, slowly increase,
reaching a shallow maximum at the same time of the current density, and then decline slowly. 
The initial growth is due to the generation of a low level of compressive fluctuations.
Velocity fluctuations (not shown) behave similarly to magnetic fluctuations, with the perpendicular component declining monotonically (there is no initial growth) and the parallel component originating from compressive effects.
In the following we will show spectra at the time of the peak of the
current density $t=200~\Omega_p^{-1}$, but all the turbulent properties are stable and remain valid until the end of the simulation ($t=500~\Omega_p^{-1}$). 

\begin{figure}[t]
\begin{center}
\includegraphics [width=0.98\linewidth]{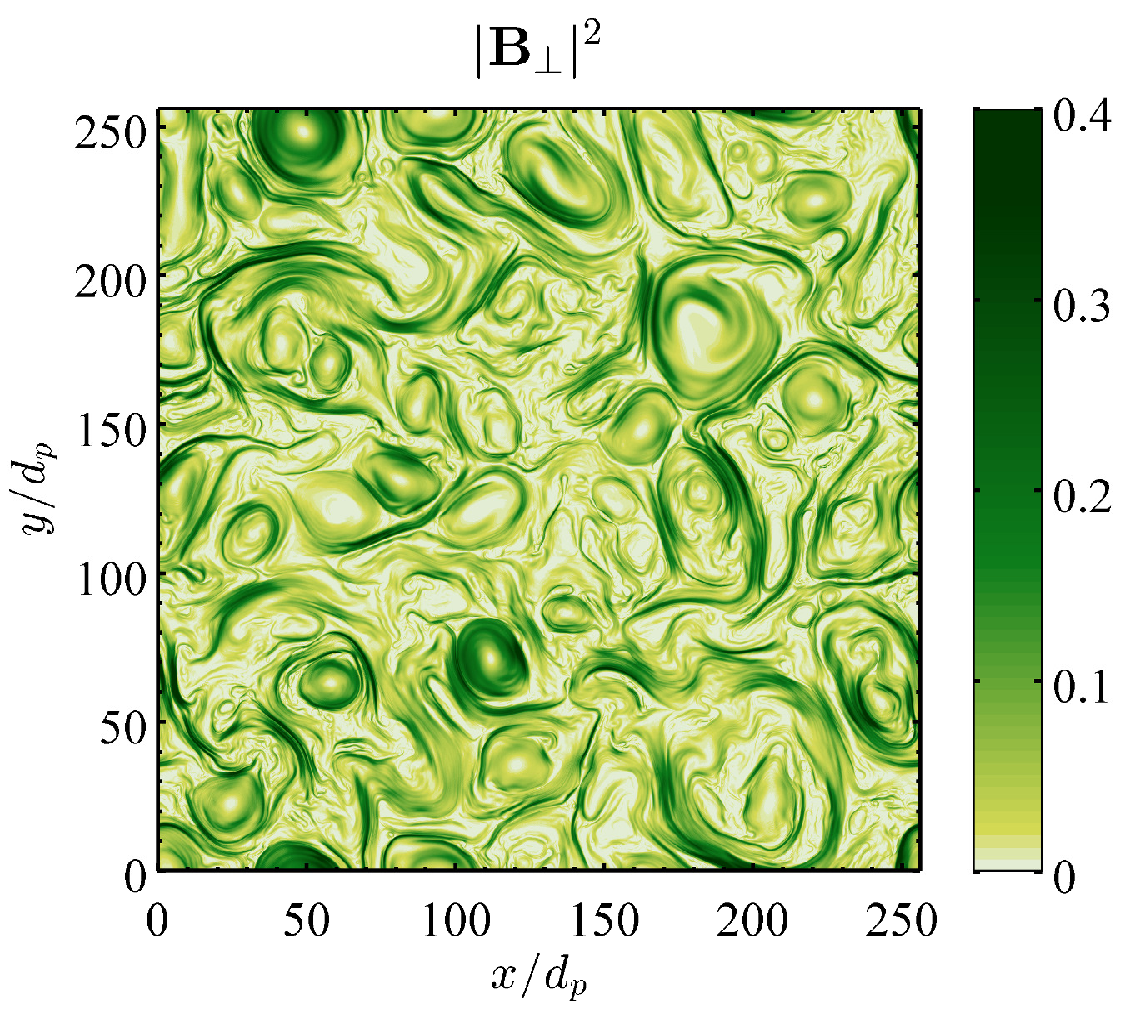}
\caption{Contour plot of the perpendicular magnetic energy at
$t=200~\Omega_p^{-1}$.}
\label{fig2}
\end{center}
\end{figure}
In Figure~\ref{fig2} we show isocontours of the perpendicular magnetic field
energy. This snapshot highlights the formation of
intense vortex-like and filamentary structures. The latter reflect the
local anisotropy of small scales fluctuations, while their random orientation
assures the statistical isotropy of the two-dimensional spectrum: we thus consider in the following only omnidirectional spectra.

\begin{figure}[t]
\begin{center}
\includegraphics [width=0.98\linewidth]{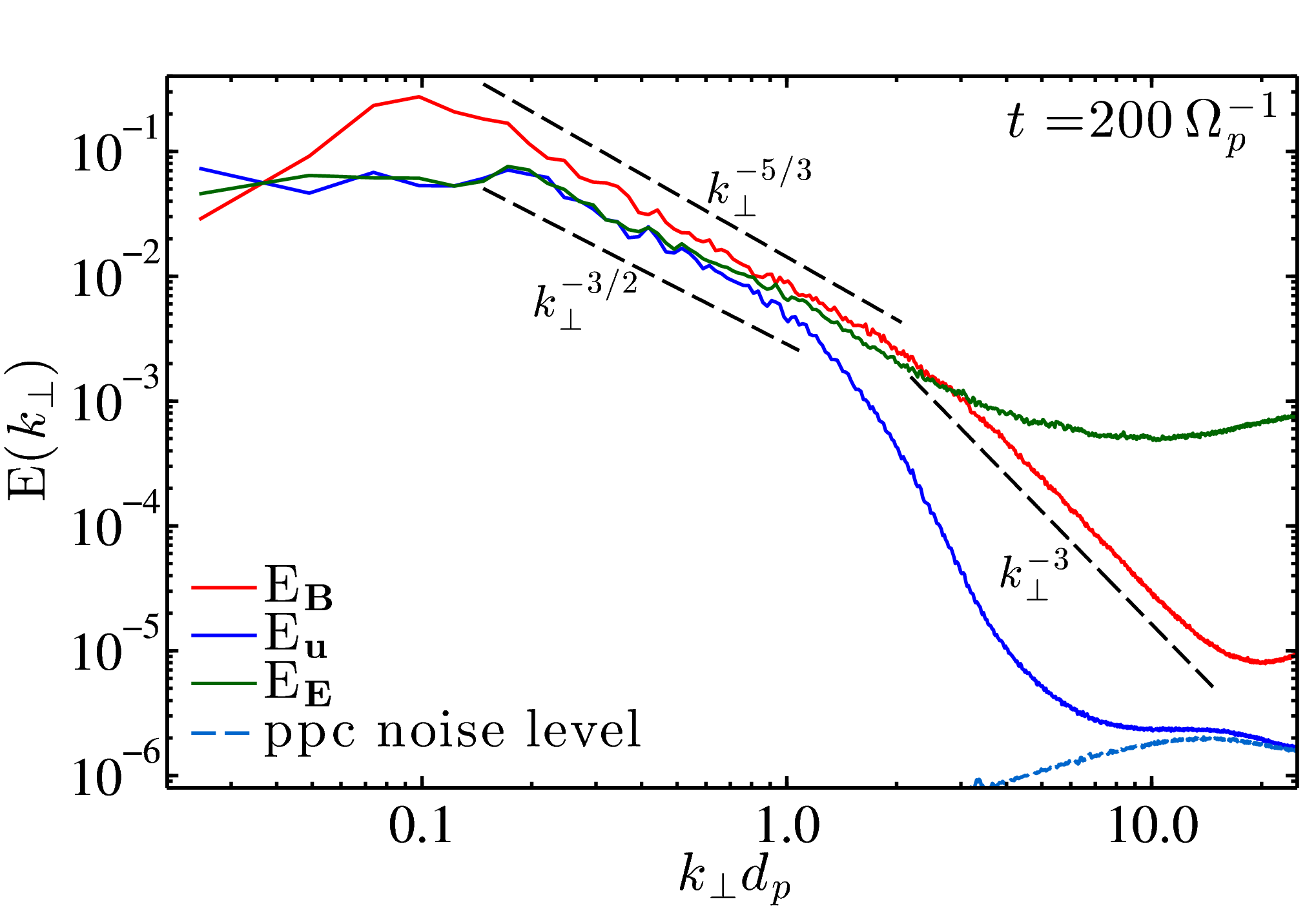}
\includegraphics [width=0.98\linewidth]{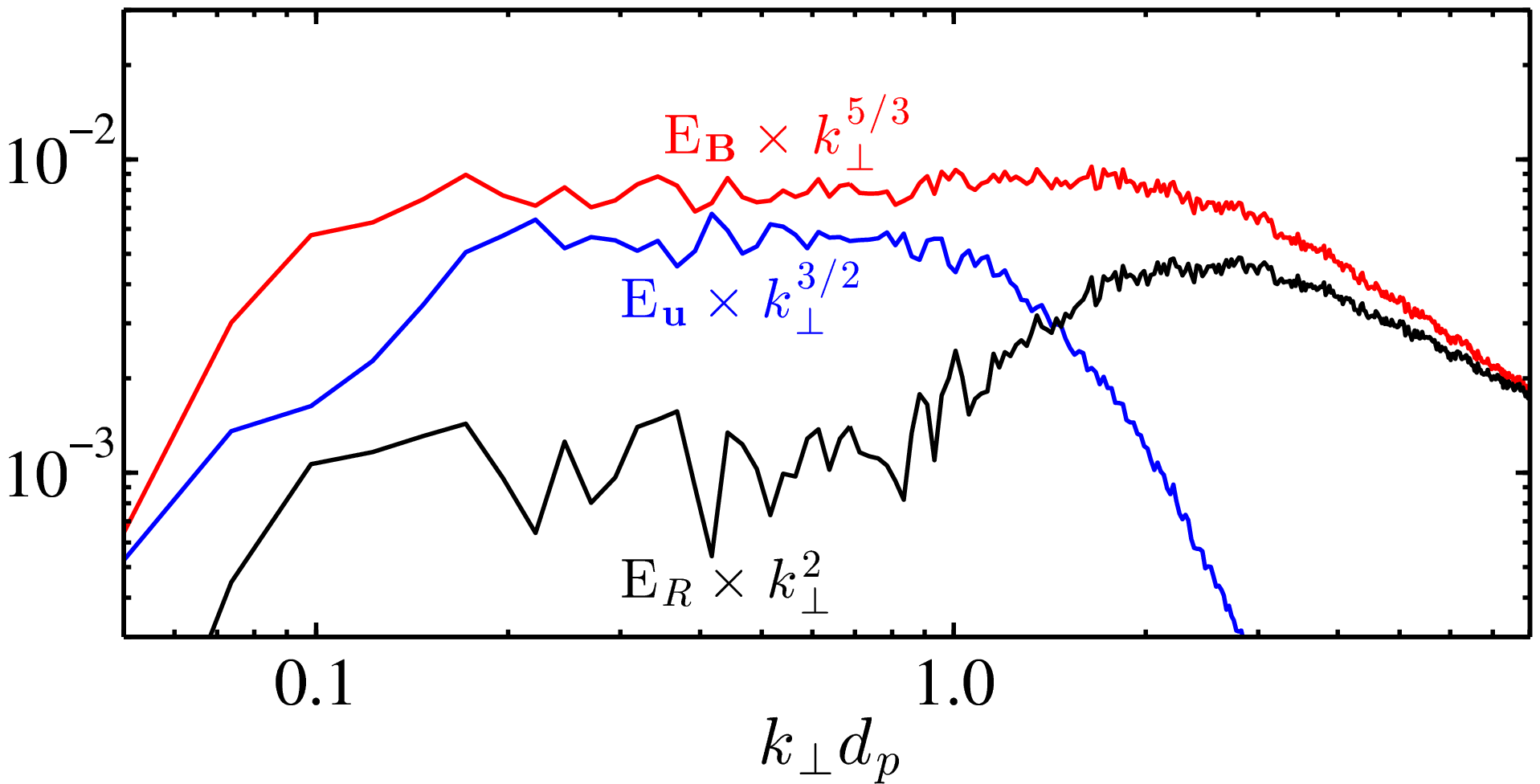}
\caption{\textit{Top panel}. Omnidirectional spectra of total magnetic (red), 
  total kinetic (blue), and total electric field (green) fluctuations versus
  perpendicular wavenumber $k_\bot$ at $t=200~\Omega_p^{-1}$.
  The spectrum of kinetic energy at $t=0$ is plotted with a dashed light-blue
  line as an indicator of ppc noise level.
  Dashed black lines are references for the corresponding
  spectral indices. 
  \textit{Bottom panel}. Magnetic (red), kinetic (blue) and residual energy
  (black) spectra compensated by $k_\bot^{5/3}$, $k_\bot^{3/2}$ and
  $k_\bot^{2}$, respectively.}
\label{fig3}
\end{center}
\end{figure}

In Figure~\ref{fig3} (top panel), we show 
the spectra of the total velocity, magnetic, and electric field.
The magnetic spectrum (red line) has a double power-law scaling, each
power-law range occupying about one decade, with a break at
$k_\bot d_p\sim2$ that separates the MHD from the sub-proton range.
The bulk velocity spectrum (blue line) also has a power-law scaling in the MHD range
but it falls off abruptly at $k_\bot d_p\sim1$, not showing any
clear power-law at higher wavenumbers. At smaller scales it reaches the ppc noise level, estimated as the level of velocity fluctuations at $t = 0$
 (light blue dashed line).
Finally the electric field spectrum (green line) follows the velocity in the
MHD range ($k_\bot d_p\lesssim 0.4$) and tends to flatten as it enters the sub-proton range ($k_\bot d_p\gtrsim 2$). 

These spectral properties are qualitatively and quantitatively in agreement with observed solar wind spectra.
In the MHD range the magnetic and kinetic spectra are
power-laws with scaling consistent with $E_B\propto k_\bot^{-5/3}$ and
$E_u\propto  k_\bot^{-3/2}$, respectively, as
can be seen in the bottom panel of Figure~\ref{fig3} where the spectra are
compensated by $k^{5/3}$ and $k^{3/2}$ respectively. In the same panel we also
plot the residual energy spectrum, $E_R=E_B-E_u$, which has a power-law scaling
over about one decade in the MHD range with a spectral index $\approx-2$ as in observations \citep{Chen_al_2013b}. 
In addition, in the sub-proton range the magnetic spectrum scales as
$E_B\propto k_\bot^{-3}$, a spectral index which is very close to the value $-2.8$ reported in observations \citep{Alexandrova_al_2009}. 
Note that the electric field spectrum is strongly coupled to the bulk velocity spectrum at MHD scales (they are basically indistinguishable for $k_\bot
d_p\lesssim 0.4$), reflecting the dominance of the ideal MHD term
($|\vect{u}\times\vect{B}|\sim B_0 \, u_\perp$) in the generalized
Ohm's law, and consistent with solar wind observations \citep{Chen_al_2011}. 
At smaller scales, it decouples from the velocity spectrum since the
Hall term ($\vect{J}\times\vect{B}/n$) and the electron pressure gradient term
($\bnabla P_e/n$) start to dominate. 

Since both other fields and derivatives enter in its
computation, $\vect{E}$ is the field that is mostly affected
by numerical effects and it's not straightforward to give a simple estimate of its noise level, as done for the velocity field.
Ultimately, we can reasonably claim that the shallower slope of its spectrum for 
$2 \lesssim k_\bot d_p \lesssim 7$ is of physical nature, while its behavior at smaller 
scales is most likely not. 
On the contrary, quantitative results for the spectra of magnetic and density fluctuations
are more robust even at larger wave numbers.
A detailed description and discussion about different sources of numerical
noise, e.g. the finite number of ppc, will be given in a companion paper \citep{Franci_al_2015}. 
For the purpose of this Letter, what matters is that such numerical noise does not affect either the 
qualitative scaling of the electric field spectrum for 
$k_\bot d_p \lesssim 7$ or the estimate of the spectral indices
of other fields up to $k_\bot d_p \sim 10$ 
(except the velocity field which is presumably affected by the noise level at $k_\bot d_p\gtrsim4$).

\begin{figure}[t]
\begin{center}
\includegraphics [width=0.98\linewidth]{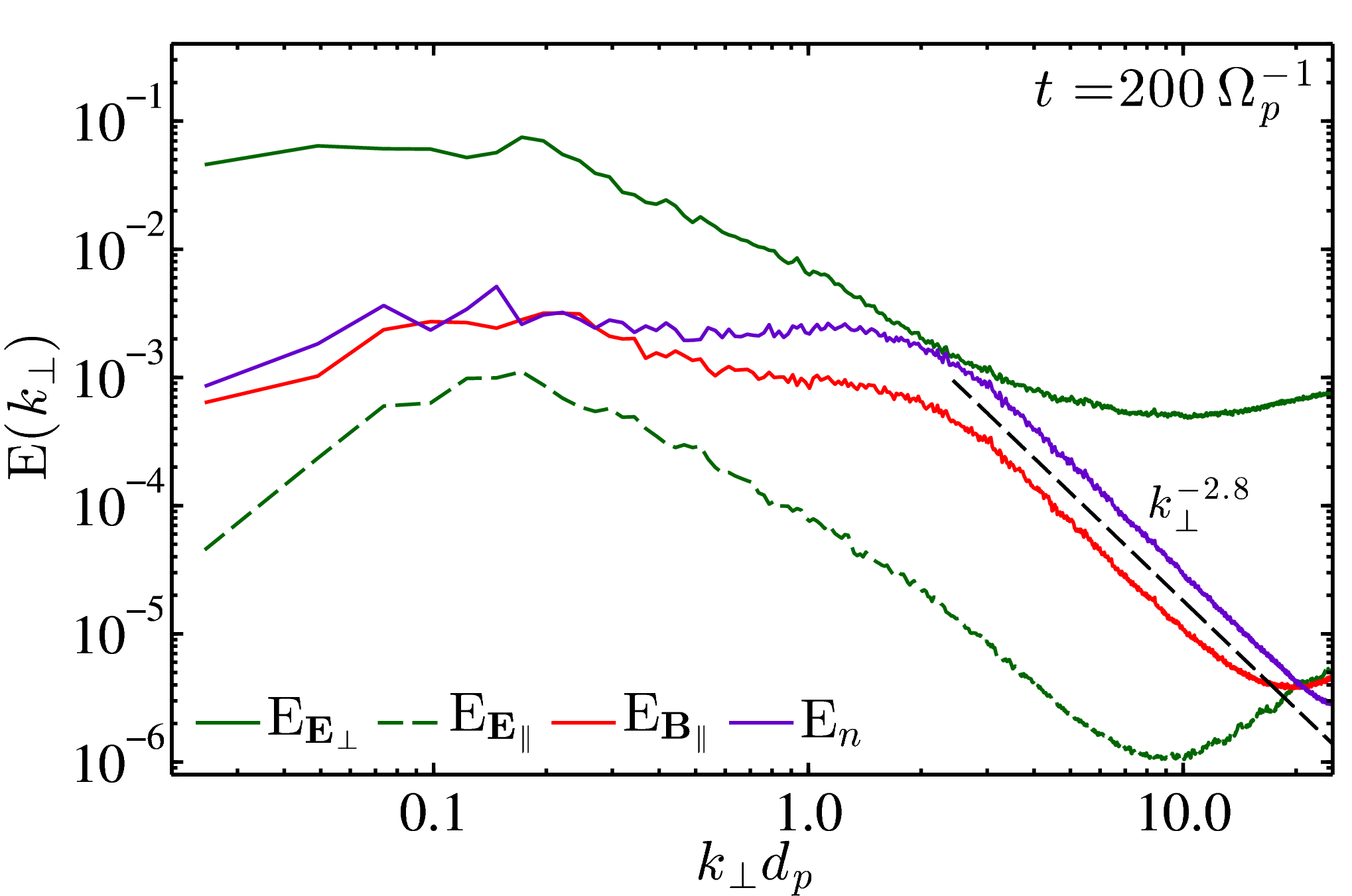}
\caption{Omnidirectional spectra of density (purple line), 
parallel magnetic field (red line), parallel and perpendicular
electric field (green dashed and solid line, respectively)
versus perpendicular wavenumber $k_\bot$ at $t=200~\Omega_p^{-1}$.}
\label{fig4}
\end{center}
\end{figure}

The transition from the MHD regime to the sub-proton regime is not only
characterized by a change in the spectral indices, but also by an increase of
energy of the parallel magnetic field and the density fluctuations relative to
other fields.
These are shown in Figure~\ref{fig4}, along with the
parallel and the perpendicular electric field spectrum. 
The density and parallel magnetic fluctuations are coupled in the whole
range of scales. In the MHD range, they have a flat spectrum that is an
order of magnitude smaller than the perpendicular electric field.
This also results in a small power in the spectrum of the total magnetic field 
intensity $E_{|B|}<E_B$ (not shown), consistently with solar wind observations \citep{Horbury_Balogh_2001}.
In the sub-proton range, $E_{B\|}$ and $E_{n}$ steepen, both having a clear
power-law scaling with index $-2.8$. By comparing Figures~\ref{fig3}-\ref{fig4} one can see that 
the parallel and perpendicular components of magnetic fluctuations become
comparable at the sub-proton scales, leading also to $E_{|B|}\sim E_B$. 
Concerning the electric field spectrum, at all scales the perpendicular component  $E_{E\bot}$ dominates by a factor $\sim100$ the parallel component $E_{E\|}$, reflecting the fact that in our configuration the leading terms of the generalized Ohm's law are linear and quadratic in the fluctuations' amplitude for $E_{E\bot}$ and  $E_{E\|}$ respectively. Note that $E_{E\bot}$ flattens at the sub-proton scales and $E_{E\|}$  steepens  in qualitative agreement with observations \citep{Mozer_Chen_2013}. 
It is hard to determine the spectral index of  $E_{E\bot}$ at sub-proton scales; a rough
estimate gives $\propto k_\bot^{-0.8}$, consistent with $\vect{E}$ being
determined by the Hall and pressure terms. In fact, retaining only the leading
order in the expression of $\vect{E}$ one gets 
$E_E \sim E_{E\bot} \propto k_\bot^2E_{B_\|,n} \sim k_\bot^{-0.8}$.

\begin{figure}[t]
\begin{center}
\includegraphics [width=0.98\linewidth]{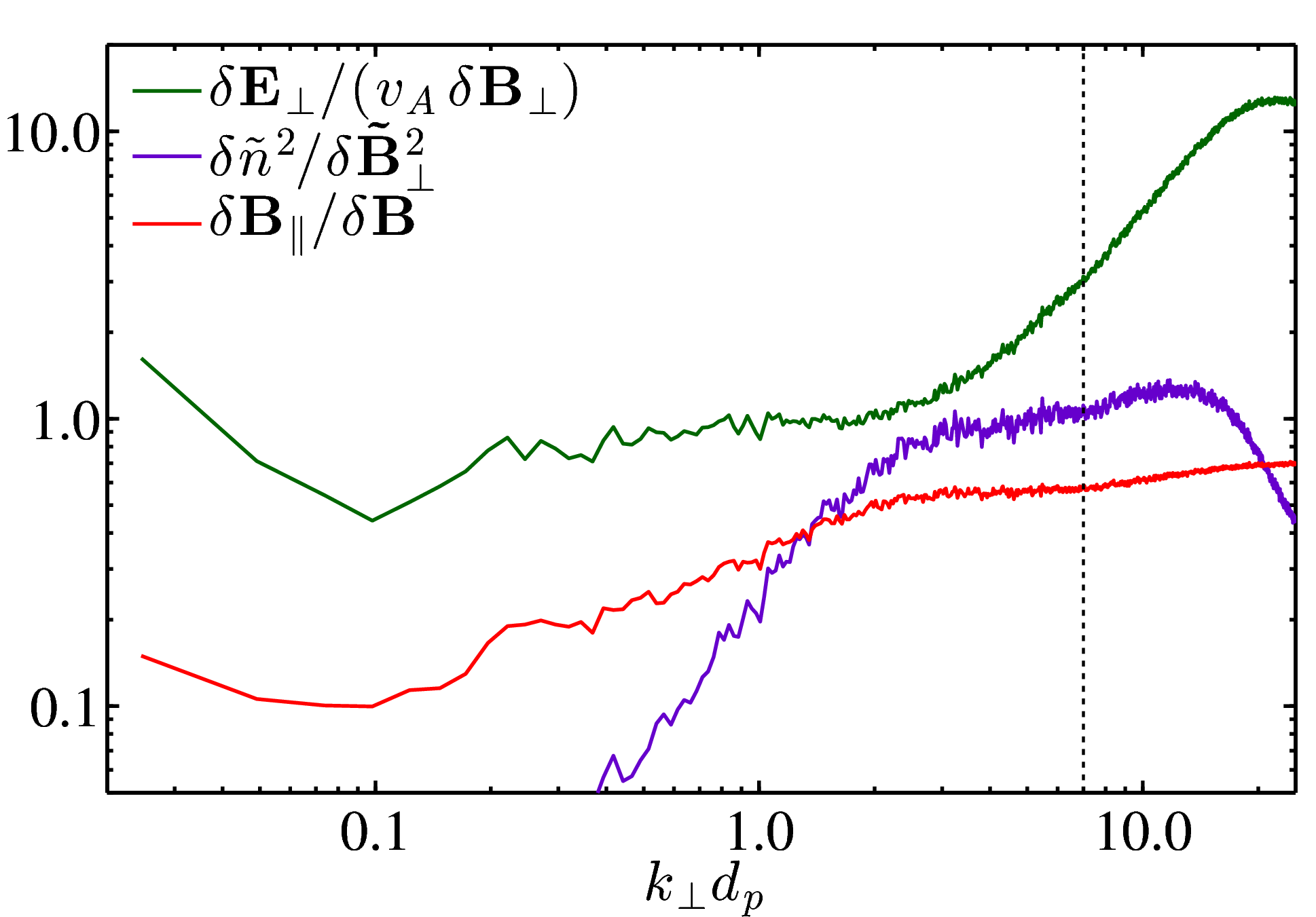}
\caption{Ratio of perpendicular electric field to perpendicular
magnetic field (green), ratio between normalized density and perpendicular magnetic fluctuations 
(purple, see text for the normalizations), and ratio of parallel to total
magnetic field fluctuations (magnetic compressibility, red). 
The numerical noise affects the ratios for $k_\bot d_p \gtrsim 7$ (vertical
dotted line).}
\label{fig5}
\end{center}
\end{figure}

We can further compare our results with observations considering three non-dimensional ratios involving density, magnetic and
electric field fluctuations shown in Figure~\ref{fig5}.
Consider first the magnetic compressibility, the ratio of parallel to
total magnetic fluctuations (red line). It is negligible in the
MHD range, increases while approaching the sub-proton scales, and finally
saturates to a level $\delta B_{||}/\delta B\sim0.5$. Thus, magnetic
fluctuations have mainly perpendicular components in the MHD range but tend to become isotropic at small scales, 
approaching a value $\delta B_{||}^2\sim\delta B_\bot^2/3$, which is within the range ($\sim0.2\div0.5$)
measured in the solar wind at spacecraft frequencies larger then $1~\mathrm{Hz}$ \citep{Kiyani_al_2013}.
This is also in very good agreement with the level of magnetic compressibility expected for kinetic Alfv\'en wave turbulence for the parameters adopted in our simulation \citep[e.g.][]{Boldyrev_al_2013}. 

The purple line in Figure~\ref{fig5} shows the ratio of normalized density
fluctuations over normalized perpendicular magnetic fluctuations,
$\delta\tilde{n}/\delta\tilde{B_\bot}$, where
$\delta\tilde{B}_\bot=\delta B_\bot/B_0$ and
$\delta\tilde{n}=\Gamma\delta n/n_o$ respectively, and $\Gamma$ ($3/4$ in our simulation) is a
non-dimensional kinetic normalization that depends on
$T_p,T_e,\beta_p,v_A$ \citep{Schekochihin_al_2009,Boldyrev_al_2013}.
With this normalization $\delta B_\perp$ and $\delta n$ are expected to have the same amplitude for kinetic Alfv\'enic fluctuations. Indeed 
$\delta\tilde{n}/\delta\tilde{B_\bot}$  increases and then saturates at a value
$\sim 1$ at sub-proton scales. 
Note that the plateau and its value $\sim1$ are consistent with observations (on
average $\delta\tilde{n}/\delta\tilde{B}_\bot=0.75$, cf. \citealt{Chen_al_2013}).

Finally, we plot the 
 ratio between the perpendicular electric
fluctuations (normalized by the Alfv\'en speed)
and the
perpendicular magnetic fluctuations (green line). Similarly to the
observed frequency spectra in the solar wind frame \citep{Bale_al_2005}, this 
ratio is about $1$ in the MHD
range, where the MHD term ($\vect{u}\times\vect{B}$) dominate. At $k_\bot
d_p\sim1$ the ratio increases reflecting the role of the
Hall term ($\vect{J}\times\vect{B}/n$) and the pressure gradient term ($\bnabla P_e/n$) 
in the generalized Ohm's law.

\section{Conclusion}
In this Letter we show that hybrid 2D large-scale, high-resolution simulations of turbulence are able to reproduce
\textit{simultaneously} several aspects of the MHD range and of the sub-proton range of solar wind spectra.
  
Two noticeable examples are given by the spectra of the magnetic field and of the electric field. The former displays a clear
double power-law scaling, with spectral indices $-5/3$ and $-3$ in the MHD and
sub-proton range respectively, separated by a smooth break at $k_\bot
d_p\sim2$. The electric field spectrum also shows a
change in the spectral properties at about the same scales, being coupled to
velocity fluctuations in the MHD range, and becomes shallower at sub-proton scale.
It is also worth noting that in the MHD range we found the scaling observed in
the solar wind for the magnetic, kinetic, and residual energy spectra (respectively $-5/3,~-3/2$, and $-2$). 
To our knowledge this is the first time that these spectral
indices are obtained for turbulence with vanishing correlation between magnetic
and velocity fields. DNS of incompressible MHD usually capture only the scaling of the residual energy and the total energy \citep{Muller_Grappin_2005} while
Reduced MHD fails in reproducing velocity and kinetic spectral indices
\citep{Chen_al_2011a} or requires special driving \citep{Boldyrev_al_2011}. This may indicate
that it is necessary to go beyond the incompressible MHD approximation even
in the inertial range. Further work is needed to test this possibility,
extending the analysis to a full 3D simulation.

In the sub-proton scales we found an increase in
magnetic compressibility and a strong coupling between
density and the parallel component of magnetic fluctuations - both having the
same spectral index of $-2.8$ - with the main cascade of $E_{B_\perp}$ driven from the MHD scales.
All these spectral indices match or are consistent with observations.
The only relevant discrepancies are the flat spectra (slope $\sim0$) of
parallel magnetic fluctuations and density fluctuations in the MHD range. 
In the solar wind they have a spectral index $-5/3$
\citep[e.g.][]{Chen_al_2012}.
This aspect is not fully captured by our simulations probably because of the
limited compressibility imposed by the 2D dynamics and/or by the value of the proton $\beta$. Note however that this does not prevent the full development of a compressible cascade at kinetic scales, in good agreement with observations.

Properties shown in Figure~\ref{fig5} are consistent with the
turbulence at sub-proton scales being ruled by fluctuations with properties of kinetic Alfv\'en waves. 
However, note that the level of magnetic and gas compressibility expected for this regime follows from more general properties of the thermodynamical state assumed for the plasma ($\beta_e$, $\beta_p$ and ion-electron temperature ratio), which govern the couplings between the different fields $B$, $E$ and $n$ via the generalized Ohm's law.
In the low-frequency regime (i.e. below the whistler range), the
ratios $\delta\tilde{n}^2/\delta\vect{B}_\bot^2$,~$\delta\vect{B}_\|/\delta\vect{B}$ are not expected to
depend on $k$ \citep{Boldyrev_al_2013} since they 
do not rely on the specific dispersion relation of the fluctuations.
In this sense, the
plateaus at $2\lesssim k_\bot\lesssim 7$ in Figure~\ref{fig5} represent a more general and likely universal manifestation of low-frequency turbulence at
kinetic scale, and this is how we intend to present them here.

As a concluding remark, we stress that our simulation implements a finite
resistivity to assure a source of damping at small scales for the magnetic fluctuations, and thus to prevent energy accumulation and the consequent artificial flattening of the spectrum. 
Although a more detailed and quantitative analysis of the related effects will be given in a forthcoming paper \citep{Franci_al_2015},
we anticipate that the values of resistivity and  the number of ppc affect the ion heating properties.

\textit{Acknowledgments} 
This project has received funding from the European Union's Seventh
Framework Programme for research, technological development and demonstration
under grant agreement No. 284515 (SHOCK). Website: project-shock.eu/home/. AV
acknowledges the Interuniversity Attraction Poles Programme initiated by the Belgian Science Policy Office (IAP P7/08 CHARM). LM was funded by STFC grant ST/K001051/1. 
PH acknowledges GACR grant 15-10057S.
HPC resources were provided by CINECA (grant 2014 HP10CLF0ZB and
HP10CNMQX2M). We warmly thank Frank L\"offler for providing HPC resources through the Louisiana State University (allocation hpc$\_$hyrel14).

\end{document}